\documentclass[a4paper]{jpconf}
\usepackage{graphicx}
\usepackage{amsmath}
\usepackage{amssymb}
\usepackage{latexsym}
\usepackage{feynmf}
\usepackage{booktabs}
\usepackage{wrapfig}
\usepackage{citesort}

\graphicspath{{images/}}
\DeclareGraphicsExtensions{.pdf}
\newcommand{\ALL}{\mathcal{A}_{\text{LL}}}
\newcommand{\aLL}{\hat{a}_{\text{LL}}}

\newcommand{\pT}{p_\text{T}}
\newcommand{\ET}{E_\text{T}}
\newcommand{\Mjj}{M_{\text{jj}}}

\newcommand{\C}{{\cal C}}
\renewcommand{\L}{{\cal L}}
\newcommand{\dt}{\text{d}t}

\newcommand{\pthat}{\hat{p}_\text{T}}

\newcommand{\GeV}{\text{GeV}}

\begin{fmffile}{fgraph}
\fmfcmd{%
vardef middir(expr p, ang) =
 dir(angle direction length(p)/2 of p + ang)
enddef;
style_def proton expr p =
 save mp, sp; pair mp; path sp;
 mp = point length(p)/2 of p;
 sp = p shifted -mp scaled 1.05 shifted mp;
 cdraw sp shifted (2thick*middir(p, 90));
 cdraw sp shifted (2thick*middir(p,-90));
 draw_fermion p
enddef;}

\begin{document}
\title{Dijet Cross Section \\ and Longitudinal Double Spin Asymmetry
\\ Measurements in Polarized Proton-proton Collisions \\ at $\sqrt{s}=$200 GeV
at STAR}

\author{Tai Sakuma$^1$ and Matthew Walker$^2$ for the STAR collaboration}

\address{$^1$Texas A\&M University,
Department of Physics and Astronomy\\
4242 TAMU, College Station, TX 77843}

\address{$^2$Massachusetts Institute of Technology,
Laboratory for Nuclear Science,
Cambridge, MA 02139}

\ead{sakuma@bnl.gov}

\begin{abstract}
These proceedings show the preliminary results of the dijet cross
sections and the dijet longitudinal double spin asymmetries $\ALL$ in
polarized proton-proton collisions at $\sqrt{s} = 200$ GeV at the
mid-rapidity $|\eta| \le 0.8$. The integrated luminosity of 5.39
pb$^{-1}$ collected during RHIC Run-6 was used in the measurements. The
preliminary results are presented as functions of the dijet invariant
mass $\Mjj$. The dijet cross sections are in agreement with
next-to-leading-order pQCD predictions. The $\ALL$ is compared with
theoretical predictions based on various parameterizations of polarized
parton distributions of the proton. Projected precision of data analyzed to date from Run-9 are shown.
\end{abstract}

\section{Introduction}

The jet production rate in \textit{polarized proton-proton collisions}
is sensitive to the \textit{polarized gluon distribution} $\Delta g(x,
Q^2)$ of the proton. The polarized gluon distribution is of particular
interest in the proton spin physics because the first moment of this
distribution is the fraction of the proton spin carried by the gluon
spin, $\Delta G$.

The $\Delta G$ is one of the four terms in the decomposition of the
proton spin in the infinite momentum frame:
\[
\frac{1}{2}=\frac{1}{2}\Sigma+\Delta G + L_q + L_g,
\]
where $\Sigma$, $L_q$, and $L_g$ are the contributions from quark spin,
quark orbital motion, and gluon orbital motion. The quark spin
contribution $\Sigma$ has been measured using \textit{polarized deep
inelastic scattering} (pDIS) data combined with neutron and hyperon
$\beta$ decay data \cite{emc1988}.

One of the primary objects of the spin physics program at RHIC
(RHIC-Spin) is to determine $\Delta G$ by using polarized proton-proton
collisions. An advantage of using proton-proton collisions is that
gluons participate in high-$\pT$ events at the leading order. However,
it is challenging to determine the kinematics of parton-level
interactions, which are desirable quantities to determine in an
experimental study of the structure of the proton. 

In order to determine the kinematics of leading-order parton-level
interactions in proton-proton collisions, the momenta of both outgoing
partons are needed:
\[
x_1 = \frac{\pthat}{\sqrt{s}}(e^{+y_3} + e^{+y_4}), \hspace{3em}
x_2 = \frac{\pthat}{\sqrt{s}}(e^{-y_3} + e^{-y_4}),
\]
where the subscripts 1, 2 indicate the incoming partons of the hard
interactions, 3, 4 indicate the outgoing partons. The momenta of
outgoing partons can be estimated by observing two final state objects in
events such as dijets and photon-jets. Furthermore, the invariant mass and average pseudorapidity of the two objects are sensitive to the kinematics according to:
\[
M_{jj} = \sqrt{x_1x_2s},\hspace{3em}
\frac{\eta_3 + \eta_4}{2} = \frac{1}{2}\ln{\frac{x_1}{x_2}}.
\]

These proceedings show the preliminary results of the
\textit{longitudinal double spin asymmetries} $\ALL$ of the dijet
production as a function of the dijet invariant mass $\Mjj$:
\[
\ALL = \frac{\sigma^{\uparrow\uparrow} - \sigma^{\uparrow\downarrow}}{\sigma^{\uparrow\uparrow} + \sigma^{\uparrow\downarrow}}.
\]
The arrows ($\uparrow\downarrow$) indicate the orientations of the helicities of the colliding protons. $\ALL$ is sensitive to the
polarized parton distributions. In fact, in the framework of
\textit{QCD factorization}, $\ALL$ can be written as follows:
\[
\ALL = \frac{\displaystyle
 \sum_{i,j}\int\text{d}x_1\int\text{d}x_2\Delta f_i(x_1, Q^2)\Delta
 f_j(x_2, Q^2)\aLL\hat{\sigma}(\cos\theta^\ast)}{\displaystyle
 \sum_{i,j} \int \text{d}x_1 \int \text{d}x_2 f_i(x_1, Q^2) f_j(x_2,
 Q^2) \hat{\sigma}(\cos\theta^\ast)},
\]
where $\Delta f_{i}$ is the polarized distribution of the parton $i$,
$f_{i}$ is the unpolarized one, $\hat{\sigma}$ is the parton-level cross
section, $\aLL$ is the parton-level longitudinally double spin
asymmetries, and $i$ and $j$ run over quark flavors and gluons.

The proceedings also show the preliminary results of the dijet cross
sections as a function of the dijet invariant mass $\Mjj$. The cross
sections are measured to support the theoretical framework which relates
polarized parton distributions and measured $\ALL$ by showing the
unpolarized equivalent of the same framework can quantitatively relate
the unpolarized parton distributions and measured cross sections.



\section{STAR Detector}

\begin{figure*}
\centering
\includegraphics[scale=0.35]{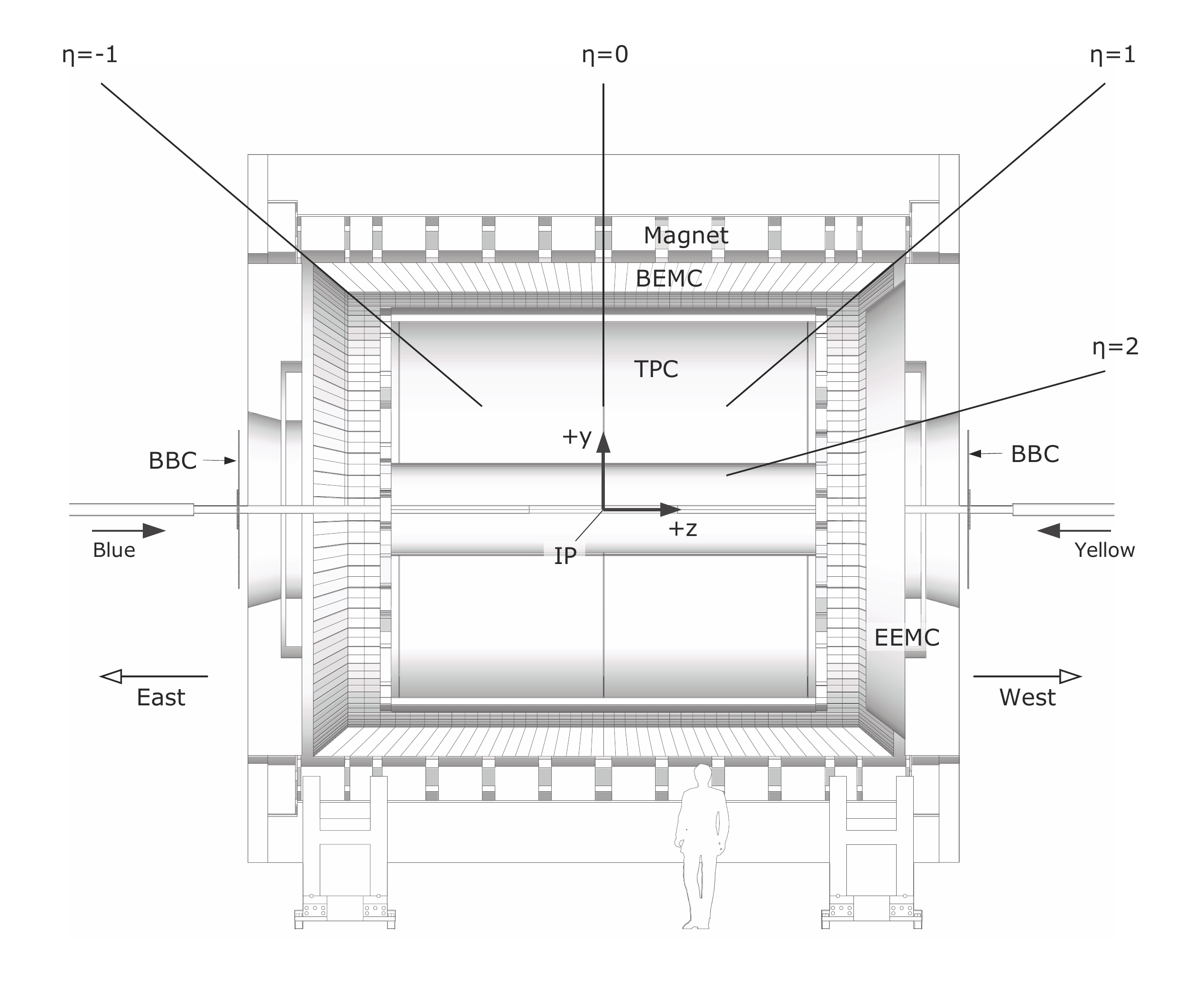}
 \caption{The STAR detector}
 \label{155900_24Aug09}
\end{figure*}

STAR, \textit{the Solenoidal Tracker At RHIC}, was built to measure wide
varieties of nuclear interactions in high energy heavy ion collisions
and polarized proton collisions \cite{Ackermann:2002ad}. Figure
\ref{155900_24Aug09} shows the cross sectional view of the STAR
detector. The detector subsystems particularly relevant to the
measurements presented in these proceedings are the \textit{Time
Projection Chamber} (TPC), the \textit{Barrel Electro\-magnetic
Calorimeter} (BEMC), and the \textit{Beam-Beam Counters} (BBC).

The \textit{Time Projection Chamber} (TPC) \cite{Anderson:2003ur} is the
primary tracking system of the STAR detector. It has a cylindrical shape
operated within a solenoidal magnetic field of 0.5 T and provides the
momentum measurements over a range from 100 $\text{MeV}$ to 30
$\text{GeV}$. Its acceptance is $|\eta| < 1.8$ with full azimuth.

The \textit{Barrel Electromagnetic Calorimeter} (BEMC) is the primary
calorimeter at mid-rapidity \cite{Beddo:2002zx}. It is a cylindrical
annulus which surrounds the TPC. The BEMC covers $|\eta| < 1$ with full
azimuth and has a depth of about twenty radiation lengths ($20 X_0$) at
$\eta = 0$. The BEMC has 4,800 towers in total and each tower covers
$\Delta\eta\times\Delta\varphi = 0.05 \times 0.05$.

The \textit{Beam-Beam Counters} (BBC) \cite{Adams:2003kv} are
scintillator annuli of hexagonal tiles. Their acceptance is
approximately $3.3 < |\eta| < 5.0$. The BBCs are used to trigger
\textit{minimum bias} (MINB) events. The MINB condition is a coincidence
between the east BBC and the west BBC. The cross section of the MINB
events is $\sigma_\text{MB} = 26.1\pm 2.0 \text{ mb}$
\cite{Drees:2001kn}. The MC simulation estimates that 87\% $\pm$ 8\% of
non-singly diffractive collisions result in a MINB trigger
\cite{Adams:2003kv}. 

\section{Event Selection}

The BJP1 (\textit{Barrel Jet Patch} 1) trigger was primarily used in the
measurements. This trigger requires a minimum transverse energy $\ET$
deposit in a \textit{patch} of calorimeter towers ($\Delta \eta \times
\Delta \varphi = 1 \times 1$) as well as the MINB condition. The $\ET$
threshold required for offline analysis was 10.8 GeV, which was above
the trigger \textit{turn-on}, to ensure high trigger efficiency.


The BBC coincidence has some allowed time difference. This difference is
measured as 4-bit values called \textit{timebin}, which roughly
corresponds to the vertex position of the events. To select events close
to the interaction point (IP), events are required to be in a specific
timebin. The vertex distribution of the events in this timebin has
approximately a Gaussian distribution with the mean -19 cm and the
standard deviation 30 cm. About 26\% of the MINB events were in this
timebin. In addition, events are required to have a reconstructed
vertex.

\section{Jet and Dijet Definition}

Jets can be defined at three different levels: the \textit{parton
level}, the \textit{hadron level}, and the \textit{detector level}. In
an experiment, jets are reconstructed at the detector level, whereas
perturbative QCD calculations predict jet productions at the parton
level. MC simulated events are used to evaluate the effects of the
transitions between different jet levels as jets can be reconstructed at
all three levels in MC simulation.

Jets are defined as collections of four-momenta selected by the
\textit{mid-point cone jet-finding algorithm} \cite{Blazey:2000qt} with
the \textit{cone radius} 0.7 and \textit{split/merge} fraction 0.5.
Four-momenta of jets are the four-vector sum of the
four-momenta that define the jets. Four-momenta that compose
detector-level jets are constructed from charged tracks in the TPC and
energy deposits in BEMC towers. Tracks are assumed to have the mass of a
charged pion (139.75 MeV), and towers are assumed to be massless. In
order to avoid measuring the same charged particles twice both in the
TPC and in the BEMC, if a track points to a BEMC tower, the energy that
a MIP would leave in the tower is subtracted from the tower energy.

In order to reject the \textit{beam-gas background}, the \textit{neutral
energy ratio} $R_\text{T}$, the fraction of jet transverse energy $\ET$
reconstructed from energy deposits in the BEMC, are required to be
smaller than particular values that depend on jet $\pT$:
$R_\text{T} < 1.0 \; (5 < p_\text{T} \le 17.31 \text{ GeV})$,
$R_\text{T} < 0.99 \; (17.31 < p_\text{T} \le 21.3 \text{ GeV})$,
$R_\text{T} < 0.97 \; (21.3 < p_\text{T} \le 26.19 \text{ GeV})$, and
$R_\text{T} < 0.90 \; (26.19 \text{ GeV} < p_\text{T})$.

Dijets are defined as the two leading-$\pT$ jets of events. Dijets are
required to contain at least one \textit{trigger jet}, a jet which
caused the BJP1 trigger. If only one jet is a trigger jet, the jet is
called the \textit{same side jet}. If both jets are trigger jets, the
jet that triggered with higher $\ET$ is the same side jet. The other jet
is called the \textit{away side jet}. Dijets are required to have
balanced $\pT$: $ 0.73 \le \pT^\text{\scriptsize
away}/\pT^\text{\scriptsize same} \le 1.1$ because $\pT$-balanced dijets
are more likely to carry momentum closer to that of parton level than do
unbalanced jets.

Asymmetric $\pT$ cuts ($\max(\pT) > 10.0\;\GeV$ and $\min(\pT) > 7.0\;\GeV$) are used because a NLO pQCD calculation
has little prediction power of dijets cross sections and $\ALL$ with
symmetric $\pT$ cuts. The $-0.8 < \eta < 0.8$ cuts are applied for the detector
acceptance. The $|\eta_3 - \eta_4| < 1.0 $ cut is necessary for both jets to be
in the acceptance at the same time. The $\Delta \varphi > 2.0$ cut is applied
to select back-to-back dijet events.


\section{MC Simulation}

\begin{figure}[h*]
\centering
\begin{minipage}{0.35\textwidth}
\centering
\includegraphics[scale=0.8]{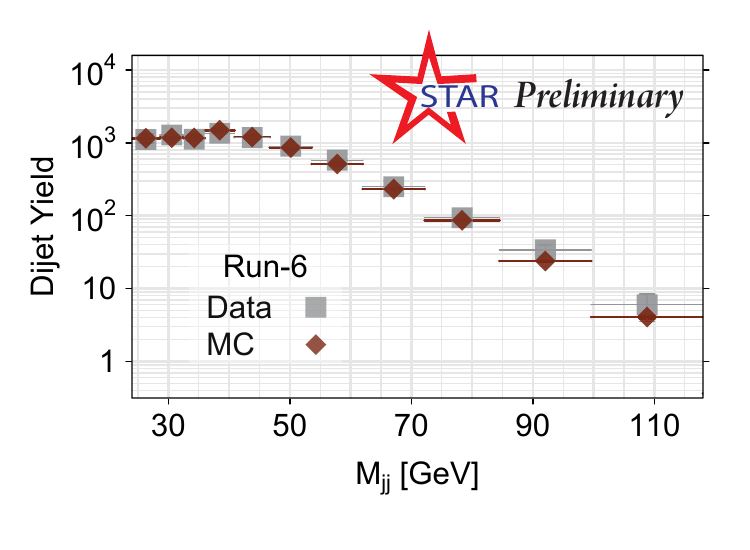}
\includegraphics[scale=0.9]{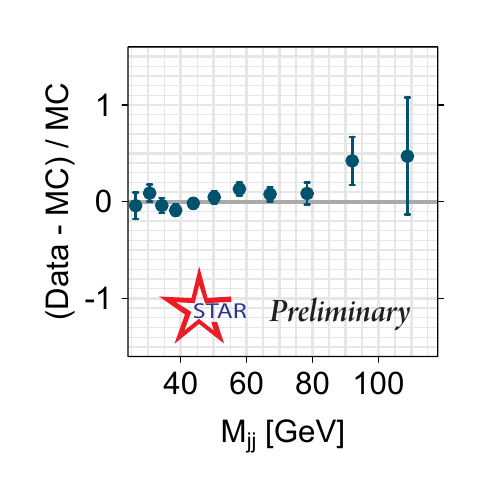}
\end{minipage}
\hspace{0.02\textwidth}
\begin{minipage}{0.58\textwidth}
\includegraphics[scale=0.7]{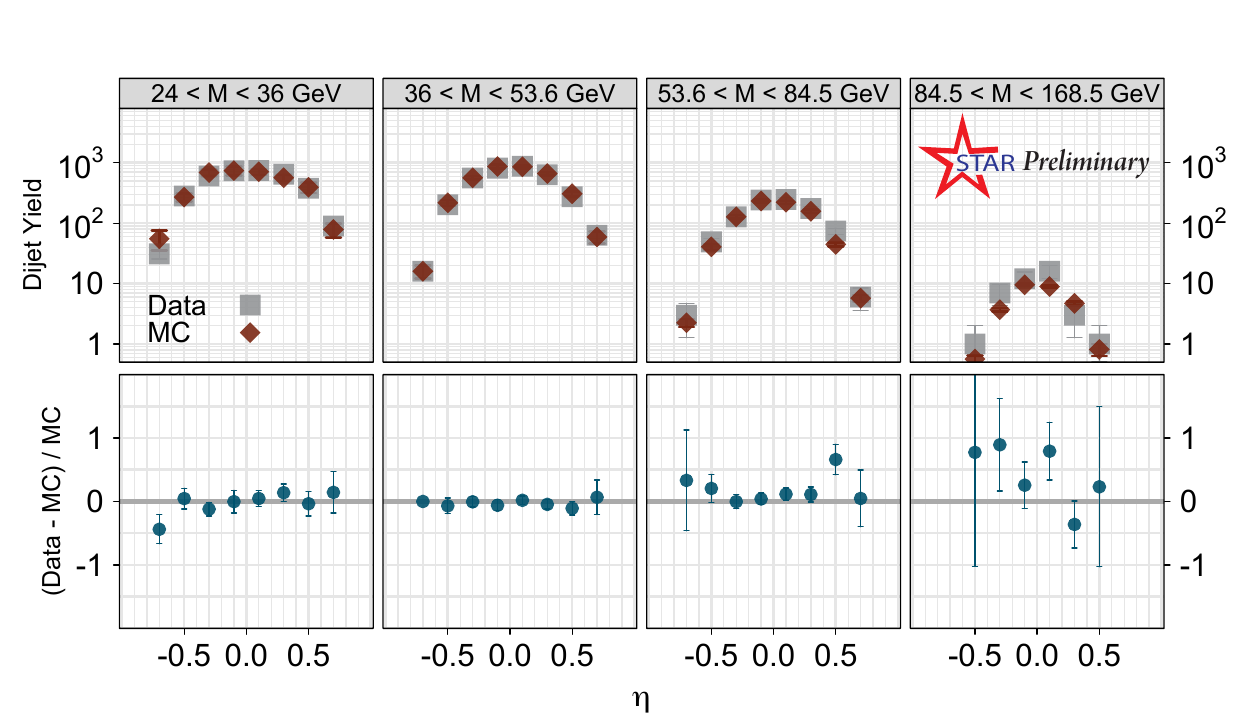}
\includegraphics[scale=0.7]{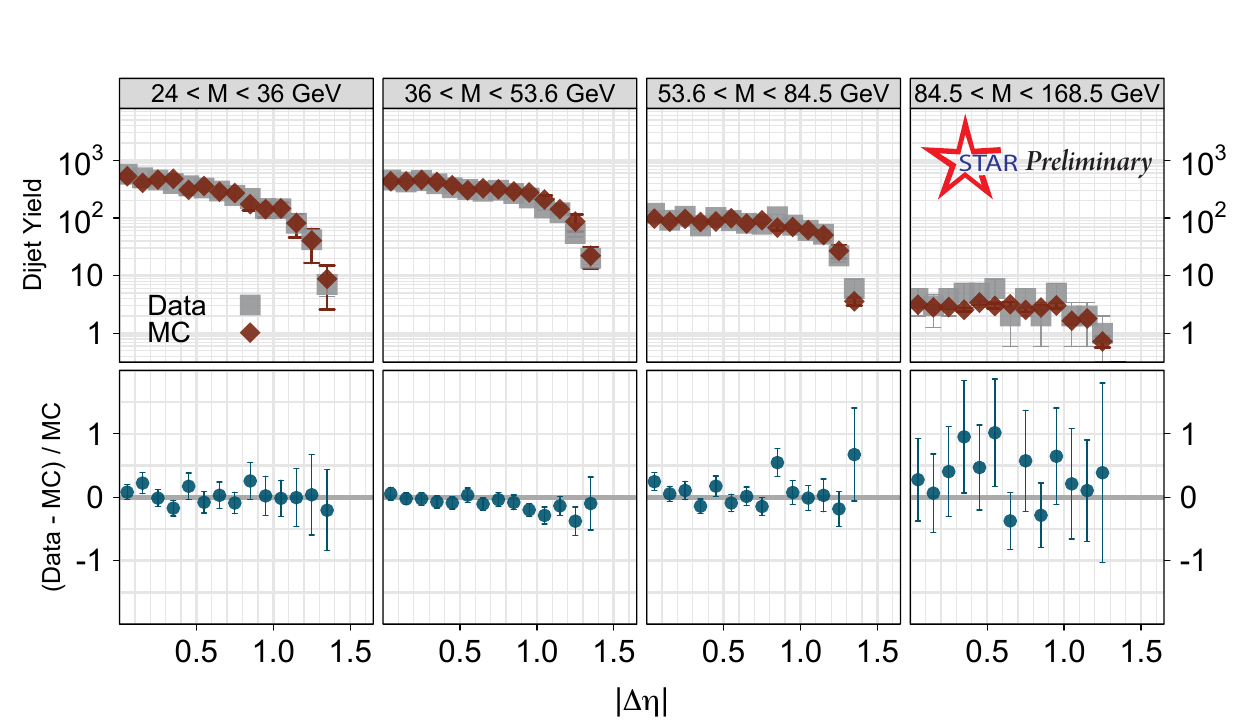}
\end{minipage}
\caption{\small The data-MC comparison of the dijet kinematic distributions.
 \textit(left) The invariant mass $\Mjj$ distributions. \textit{(top
 right)} The average pseudo-rapidity $\eta$ distributions.
 \textit{(bottom right)} The pseudo-rapidity difference $\Delta\eta$
 distributions. The MC yields are scaled so that the yield becomes the
 same as the data. }
\label{185510_14Nov10}
\end{figure}

The events are generated by the Pythia 6.410 event generator
\cite{Sjostrand:2006za} with the CTEQ5L\index{CTEQ5L} parton
distributions \cite{Lai:1999wy} using parton $p_T$ between 3 and 65 GeV. The detector responses to the events
are simulated with a GEANT3 \cite{geant3} based STAR detector simulation
program.

In the MC simulation, jets are reconstructed at all three jet levels
using the same jet finder as in the data. The detector-level jets are
defined in the same way as in the data. The hadron-level jets are
collections of final state particles in the event generator, while the
parton-level jets are composed of outgoing partons of the hard
interactions and the radiation from the outgoing partons.

The MC events well reproduce the data. Fig. \ref{185510_14Nov10} shows
the data-MC comparison of the dijet kinematic distributions: the
invariant mass $\Mjj$, the average pseudo-rapidity $\eta = (\eta_3 +
\eta_4)/2$, and the pseudo-rapidity difference $\Delta\eta = \eta_3 -
\eta_4$ distributions. 

\section{Cross Section Measurement}


The dijet cross sections are estimated for each $\Mjj$ bin with the
formula:
\[
\frac{\text{d}^3\sigma}{\text{d}\Mjj\text{d}\eta_3\text{d}\eta_4}=\frac{1}{\int {\cal
L}\text{d}t}\cdot\frac{1}{\Delta\Mjj\Delta\eta_3\Delta\eta_4}\cdot\frac{1}{\C}\cdot J.
\]
$J$ is the dijet yields at the detector level. $\C$ is the correction
factors which correct the dijet yields to the hadron level. The
correction factors $\C$ are estimated from the MC events as bin-by-bin
ratios of the dijet yields at the detector level and at the hadron level.
$\Delta\Mjj\Delta\eta_3\Delta\eta_4$ normalizes the cross sections to
per unit space volume. $\int\L\dt = 5.39 \pm 0.41\text{ pb}^{-1}$ is the
integrated luminosity measured using the BBCs.


The major systematic uncertainty is due to the uncertainty on the jet
energy scale (JES). Because the $\Mjj$ dependence is steeply decreasing,
the uncertainty of the cross section is very sensitive to systematic
uncertainty on the JES. 

The track portion of the jet energy has 5.6 \% of systematic uncertainty. To evaluate the effect of this uncertainty on the cross
sections, the cross section was reevaluated with the 5.6 \% variation of
the track portion of the jet energy.

Energy deposits in the BEMC towers have 4.8 \% of systematic uncertainty. The effect of this uncertainty was evaluated by varying
the BEMC tower energies. After the energies were varied, the offline
trigger thresholds were reapplied and the jet finding algorithm was
rerun.

Systematic uncertainty due to the correction for the pile-up and the
timebin selection are estimated to be small compared to the JES
uncertainty. The cross section has 7.6\% of correlated systematic
uncertainty due to the uncertainty of the integrated luminosity.


The dijet cross sections are calculated by next-to-leading order
perturbative QCD with the CTEQ6M \cite{Pumplin:2002vw} parton
distributions. Jets are defined by a cone jetfinding algorithm with cone radius 0.7. Both the renormalization scale and
the factorization scale are $\mu = \Mjj$. The scale uncertainty is
calculated by varying the scale from $0.5\Mjj$ to $2\Mjj$.
 
The effects of the hadronization and the underlying events were
evaluated by using the MC events. The correction factors $C_\text{HAD}$
were obtained for each $\Mjj$ bin as the ratios of the cross section at
the hadron level and at the parton level. The systematic uncertainty on
$C_\text{HAD}$ was calculated by varying the cone radius from 0.6 to
0.8.

\begin{figure}[h*]
\centering
\includegraphics[width=0.45\textwidth]{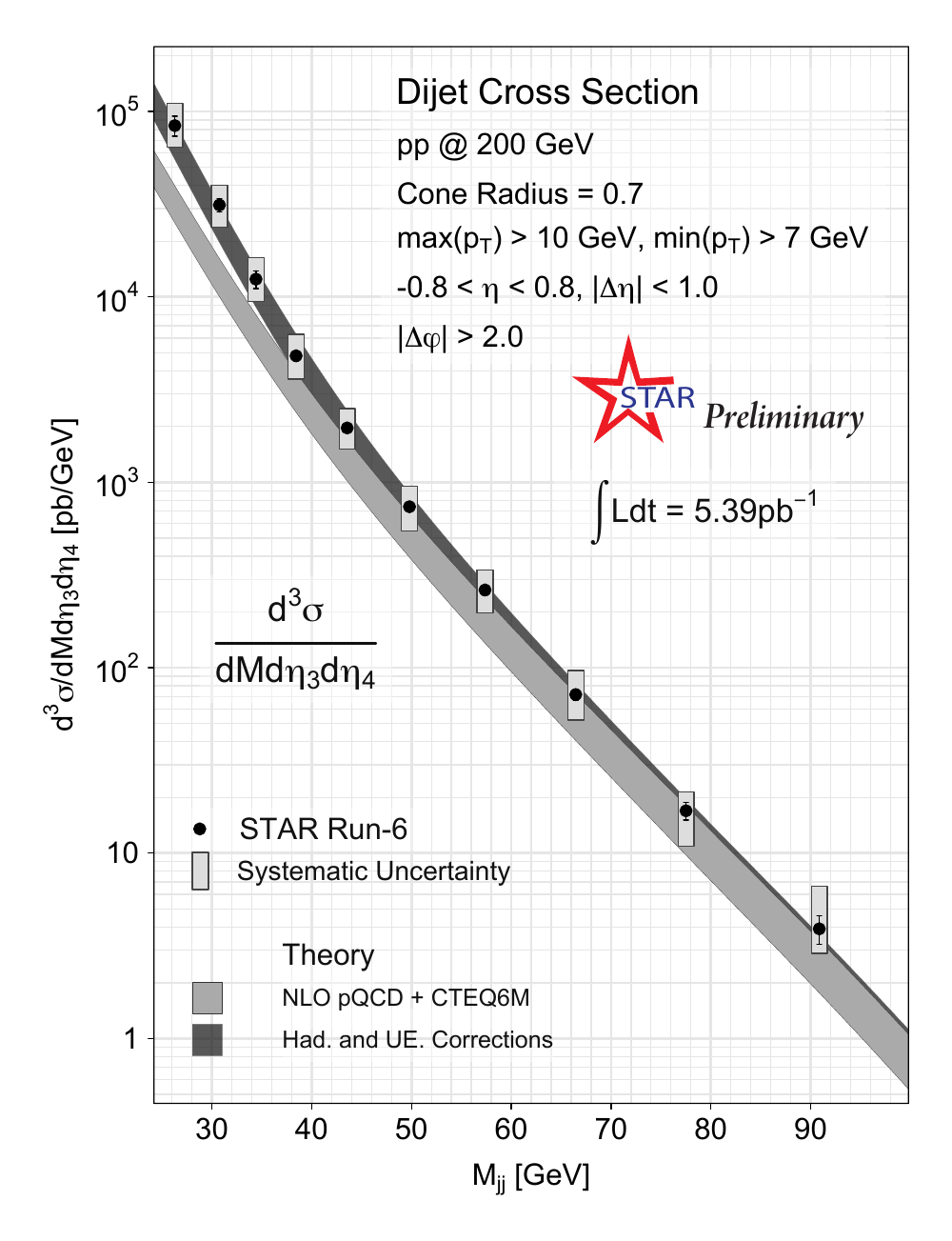}
\hspace{2pc}
\includegraphics[width=0.45\textwidth]{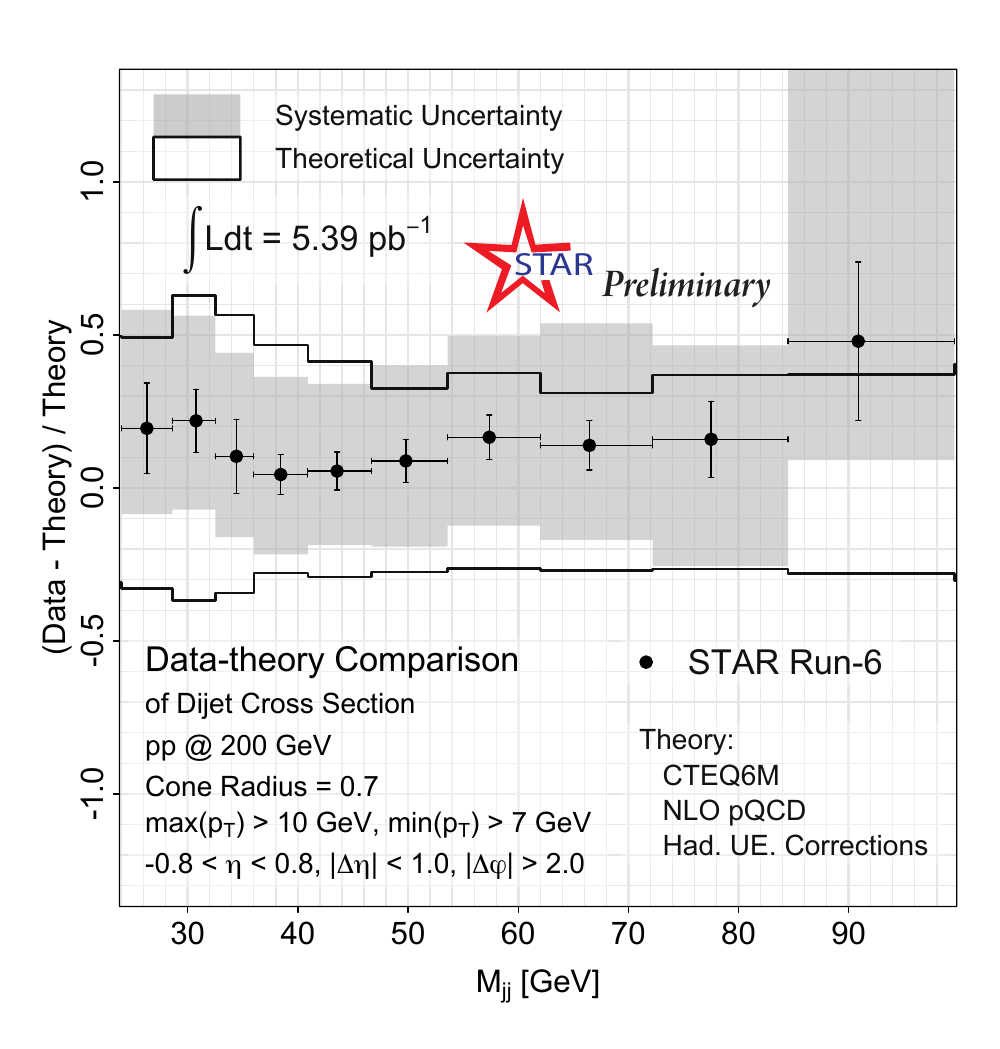}
\caption{\small\textit{(left)} The dijet cross sections compared to
theoretical predictions. The systematic uncertainty does not include
7.68\% of uncertainly due to the uncertainly on the integrated
luminosity. \textit{(right)} The ratios: (data - theory)/theory. The
theory includes the NLO pQCD predictions and the corrections for the
effects of hadronization and underlying events. The ratios are taken bin
by bin.} \label{201523_15Nov10}
\end{figure}

Figure \ref{201523_15Nov10} shows the preliminary results of the dijet
cross section measurements. The measured dijet cross sections are well
described by the theory. This indicates that measured $\ALL$ as well can
be interpreted in the same theory and suggests ways to constrain the
polarized gluon distributions from dijet $\ALL$.

\section{Longitudinal Double Spin Asymmetry $\ALL$ Measurement}


The dijet longitudinally double spin asymmetries $\ALL$ were measured as
the ratios of the spin sorted dijet yields with the corrections for the
relative luminosity and polarizations:
\[
\displaystyle \ALL=\frac{\sum P_{\text Y}P_{\text B}\{(N_{\uparrow\uparrow}+N_{\downarrow\downarrow})-R(N_{\uparrow\downarrow}+N_{\downarrow\uparrow})\}}{\sum P^2_{\text Y}P^2_{\text B}\{(N_{\uparrow\uparrow}+N_{\downarrow\downarrow})+R(N_{\uparrow\downarrow}+N_{\downarrow\uparrow})\}}.
\]
The arrows ($\uparrow\downarrow$) indicate the orientations of the helicities of the proton beams. The relative luminosity $R =
(\L_{\uparrow\uparrow} +
\L_{\downarrow\downarrow})/(\L_{\uparrow\downarrow} +
\L_{\downarrow\uparrow})$ was measured using the BBCs. The relative
luminosity varied between approximately 0.9 and 1.1. The polarizations
($P_{\text Y}P_{\text B}$) was measured by the pC CNI polarimeter and
the polarized H jet polarimeter \cite{Okada:2008p764}. The average
polarization for the Yellow beam and the Blue beam of RHIC were
$\overline{P_{\text Y}} = 59\%$ and $\overline{P_{\text B}} = 56\%$,
respectively. The \textit{figure-of-merit} = $P_{\text Y}^2 P_{\text B}^2 \L$ for this measurement is 0.59 pb\textsuperscript{-1}.


Four false asymmetries, which should vanish, are measured for a
systematic check of the data. Two single spin asymmetries and two
wrong-sign spin asymmetries were consistent with zero within the
statistical uncertainties.

The largest systematic uncertainty is the trigger bias, which is the
uncertainty in the changes of $\ALL$ from the parton level to the
detector level. This was evaluated using the MC events with several
polarized parton distributions which are compatible with the current
experimental data: DSSV, GRSV std, and the GRSV series with $\Delta G$
from -0.45 to 0.3 \cite{Gluck:2001p1625}, \cite{deFlorian:2008mr}. In this evaluation, in order to
calculate $\ALL$ with the unpolarized event generator Pythia, the MC
events were weighted by the products of the spin asymmetries of the
parton distributions and parton-level cross sections: $(\Delta
f_1(x_{1i}, Q_i^2)\Delta f_2(x_{2i}, Q_i^2)/f_1(x_{1i},
Q_i^2)f_2(x_{2i}, Q_i^2))\cdot\aLL(\cos\theta_i^*)$.

\begin{figure*}[h*]
\centering \includegraphics[scale=0.7]{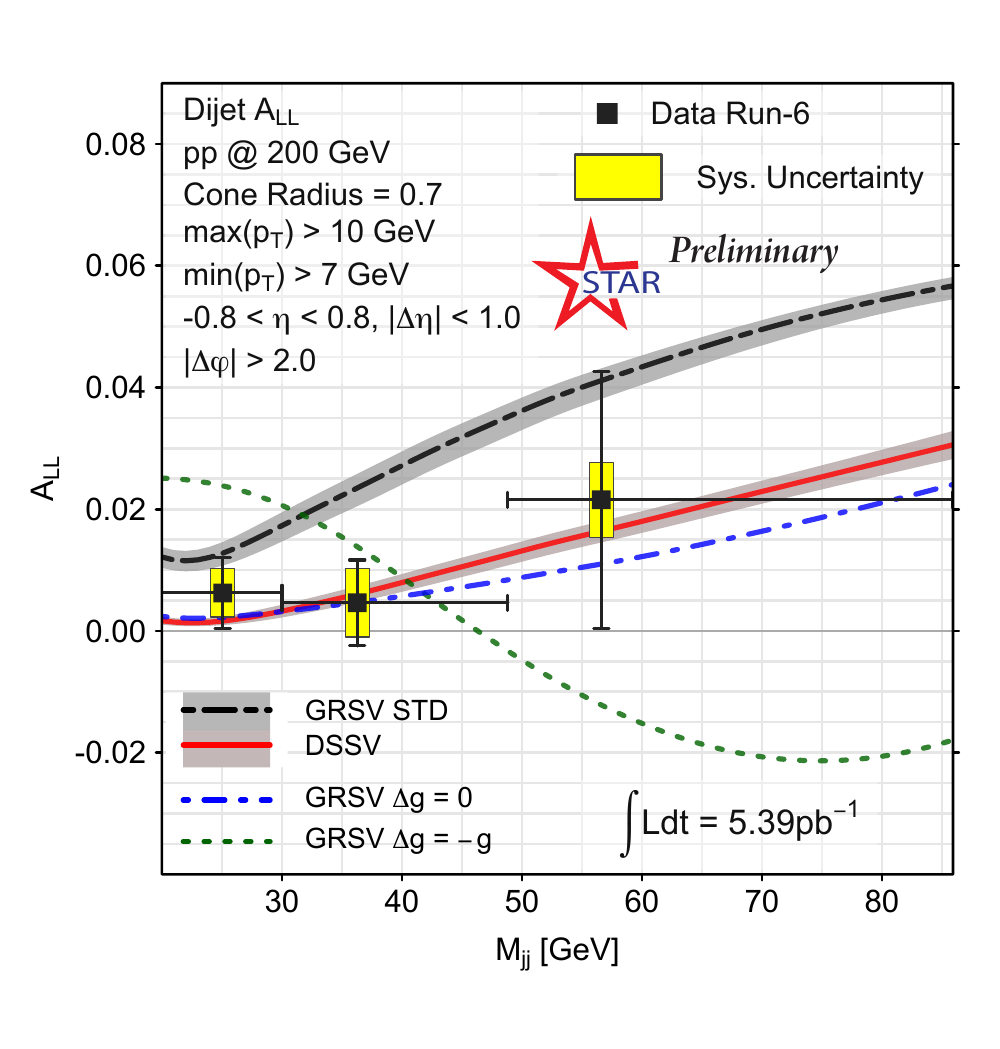}
\caption{\small The double longitudinal spin asymmetry $\ALL$ for the
dijet production as a function of dijet mass $\Mjj$. The vertical bars
on the data points indicate the size of the statistical errors. The
horizontal bars on the data points indicate the bin widths. The data
points are plotted at the mean values of $\Mjj$ of the events in the
bins. Predictions of next-to-leading perturbative QCD with various
models of the polarized gluon distributions are shown.}
\label{035228_2Dec09}
\end{figure*}

Figure \ref{035228_2Dec09} shows the preliminary results of dijet $\ALL$
measurements. It can be seen that the results are consistent with the
next-to-leading perturbative QCD prediction of the DSSV polarized
parton distributions. The results are also consistent with GRSV zero scenario, and larger values of $\Delta g$ in GRSV are disfavored.

\section{Projected Sensitivity}
Data collected during 2009 at RHIC represents a considerable increase in sensitivity to $A_{LL}$ for dijet production. Approximately 22 pb\textsuperscript{-1} were recorded with an average polarization of 59\% in both beams. Data corresponding to a figure-of-merit of 0.96 pb\textsuperscript{-1} have been processed using the TPC and BEMC to measure dijets at mid-rapidity. With the additional statistics, the data can be divided into different pseudorapidity acceptances, which provides sensitivity to the ratio of $x_1/x_2$. The invariant mass distribution for different pseudorapidity acceptances is therefore sensitive to different $x_1$, $x_2$ phase space and allows extraction of constraints on the shape of $\Delta g(x)$.

\begin{figure}
\begin{center}
\includegraphics[scale=0.6]{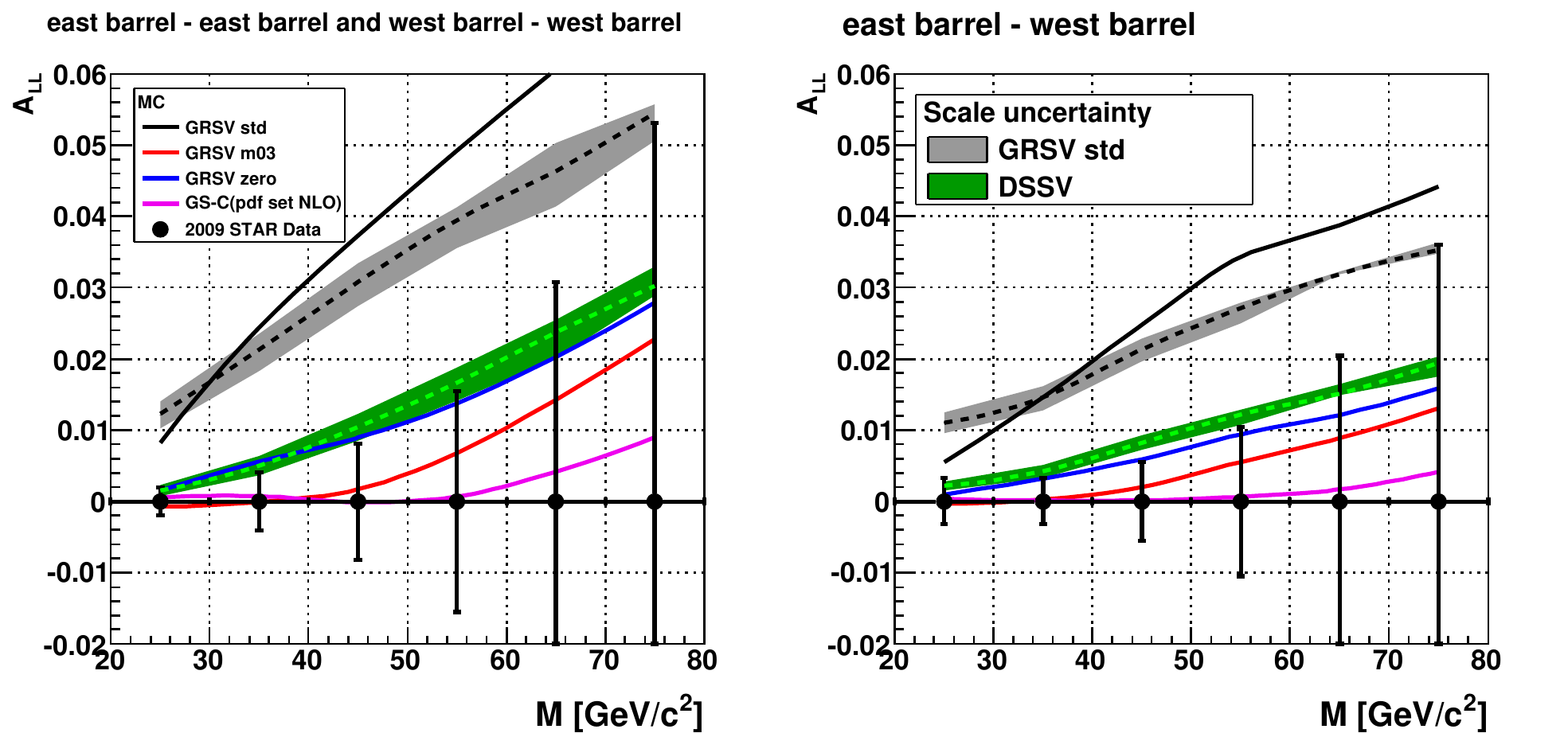}
\caption{\small\label{fig:all_ew} The statistical precision of 2009 STAR dijet data analyzed to date. The data has been divided into two pseudorapidity regions, which provide different sensitivities to the Bjorken-x phase space. The data in the left (right) panel are from when the two jets have the same (opposite) sign pseudorapidity.}
\end{center}
\end{figure}

The two interesting divisions are when the two jets have the same sign pseudorapidity and when they have opposite signs. The statistical precision of the data analyzed to date can be seen in Fig. \ref{fig:all_ew}, which is improved over comparable figure-of-merits from previous years due to improvements in trigger efficiency. An expected increase in the figure-of-merit by a factor of between 1.5 and 2.0 will provide further improvement of the sensitivity. Various studies to understand systematic uncertainties are underway.

\section{Summary}
These proceedings showed the preliminary results of the dijet cross
sections and dijet $\ALL$ in polarized proton-proton collisions at
$\sqrt{s}$ = 200 GeV from Run-6. These results agree well with pQCD calculations and provide constraints on the gluon polarization $\Delta g(x)$ in the proton. Projected precision from Run-9 data in multiple acceptances shows that STAR will be able to add new constraints to the shape of $\Delta g(x)$.

\ack 

We thank Daniel de Florian for providing tools for theoretical
calculations. We thank the RHIC Operations Group and RCF at BNL, and the
NERSC Center at LBNL for their support. This work was supported in part
by the Offices of NP and HEP within the U.S. DOE Office of Science; the
U.S. NSF; the BMBF of Germany; CNRS/IN2P3, RA, RPL, and EMN of France;
EPSRC of the United Kingdom; FAPESP of Brazil; the Russian Ministry of
Education and Science; the Ministry of Education and the NNSFC of China;
IRP and GA of the Czech Republic, FOM of the Netherlands, DAE, DST, and
CSIR of the Government of India; Swiss NSF; the Polish State Committee
for Scientific Research; SRDA of Slovakia, and the Korea Sci. \& Eng.
Foundation. Finally, we gratefully acknowledge a sponsored research
grant for the 2006 run period from Renaissance Technologies Corporation.

\section*{References}
\bibliography{dijet_spin10}

\end{fmffile}
\end{document}